\documentstyle[titlepage,12pt,psfig]{article}
\begin{document}
\title {Quantum Evolution of the Bianchi Type I Model}
\author { V.N.~Folomeev and
V.Ts.~Gurovich \thanks{email: gurovich@grav.freenet.bishkek.su}
\\
{\it Physics Institute of NAN KR,}\\
         {\it 265 a, Chui str., Bishkek, 720071,  Kyrgyz Republic.}}
\maketitle

\begin{abstract}
The behaviour of the flat anisotropic model of the Universe with a scalar
field is explored within the framework of quantum cosmology. The principal
moment of the account of an anisotropy is the presence either negative
potential barrier or positive repelling wall. In the first case occur the
above barrier reflection of the wave function of the Universe, in the second
one there is bounce off a potential wall. The further evolution of the
Universe represents an exponential inflating with fast losses of an
anisotropy and approach to the standard cosmological scenario.
\end{abstract}

\par\bigskip
\noindent {\large \bf 1. \quad Introduction}
\par\bigskip
\noindent One of the basic problems of the standard cosmological scenario is the
presence of an initial singularity. For its elimination are used the various
approaches, one of which is a quantum cosmology originated by DeWitt~[1]
more than thirty years ago. In last one the whole Universe is
described with the help of the wave function. Until recently main part of
papers was devoted to the consideration of the process of quantum creation
of the closed Universe. In this direction by number of authors [2]
was developed the scenario of the spontaneous creation of the
closed Universe from ''nothing'', where ''nothing'' means the state with
absence not only matter, but also the spacetime in classical understanding.
Thus it is necessary to note that the closed Universe has a zero total
energy and charge therefore there is no violation of any conservation laws.
At describing of process of creation of the Universe the so-called instanton
method [3] is widely used. In the last one the closure of the
Universe is a necessary requirement of the quantum creation, otherwise the
action is equal to infinity and, hence, relevant probability of the creation
is near to zero. However, recently observant data have appeared according to
which our Universe probably is open. A number of models of the quantum
nucleating of the open Universe were proposed in this connection. One of its
is the Hawking-Turok instanton [4], for which the opportunity
of an analytic continuation of the instanton solution not only in the
direction of creation of the closed Universe, but also in the direction
of the open Universe is shown. This result, being interesting on itself, is
not unexpected. For a long time it is known (see e.g. [5])
that dividing by various ways the four-dimensional de Sitter Universe on
space and time we can obtain the various de Sitter Universes: closed, flat
or open. The quantum creation of the open Universe can seem not actual in
view of the noted above infinite value of action. But here it is possible to
remember the theory of nucleation of the open Universe from bubble
[6,7]\ filled by false vacuum, when the Universe inside of this bubble
looks finite from the point of view of an external observer and open, that
is infinite, from the point of view of an inside observer. Therefore, for
an external observer the process of the instanton nucleation of the open
Universe is quite possible and the growth of the bubble size is carried out
by transferring of energy from the surrounding de Sitter space.

\par\bigskip

\noindent {\large \bf 2.\quad Basic classical equations}

\par\bigskip

\noindent We shall consider an anisotropic Bianchi type I model. For it the
synchronous form of metric can be expressed as (velocity of light is equal
1):
\begin{equation}
\label{eq1}
ds^{2}=dt^{2}-a_{1}^{2}(t)dx^{2}-a_{2}^{2}(t)dy^{2}-a_{3}^{2}(t)dz^{2},
\end{equation}
where by $a_{1},a_{2}$ and $a_{3}$ the scale factors on directions $x$, $y$,
and $z$ accordingly are designated. This model is an anisotropic
generalization of the Friedmann model with Euclidean spatial geometry. Three
scale factors $a_{1},a_{2}$ and $a_{3}$ are determined via Einstein's
equations. For convenience of realization of the analytical calculations we
can write them as follows [8] :
\begin{equation}
\label{eq2}
a_{1}=r(t)q_{1},\,\,\,a_{2}=r(t)q_{2},\,\,\,a_{3}=r(t)q_{3}\,,
\end{equation}
where $q_{1},q_{2},q_{3}$ are dimensionless variable subordinated to the
following requirements:
\begin{equation}
\label{eq3}
\prod\limits_{\alpha =1}^{3}q_{\alpha }=1,\,\,\,\sum\limits_{\alpha
=1}^{3}\left( \dot {q}_{\alpha }/q_{\alpha }\right) =0,
\end{equation}
(dot means derivative with respect to time $t$); whence follows that
$\prod\limits_{\alpha=1}^{3}a_{\alpha }=r^{3}$.

For the line element (\ref{eq1}) with the account of (\ref{eq2}) the components of the
Ricci tensor write as:
\begin{eqnarray}
\label{eq4}
-R_{0}^{0}&=&3\frac{\ddot {r}}{r}+\sum\limits_{\alpha =1}^{3}\left(
\frac{\dot {q}_{\alpha }}{q_{\alpha }}\right) ^{2},  \nonumber \\
-R_{\alpha }^{\alpha } &=&\frac{\ddot {r}}{r}+2\left( \frac{\stackrel{%
.}{r}}{r}\right) ^{2}+3\frac{\dot r}{r}\frac{\dot {q}%
_{\alpha }}{q_{\alpha }}+\left( \frac{\dot {q}_{\alpha }}{q_{\alpha }}%
\right) ^{\mbox{\bf .}}, \\
-R &=&6\left[ \frac{\ddot {r}}{r}+\left( \frac{\dot {r}}{r}%
\right) ^{2}\right] +\sum\limits_{\alpha =1}^{3}\left( \frac{\dot {q}%
_{\alpha }}{q_{\alpha }}\right) ^{2}.  \nonumber
\end{eqnarray}
Let's use the last one for obtaining (1-1) and (2-2) components of the
Einstein tensor:
\[
G_{1}^{1}=2\frac{\ddot {r}}{r}+\left( \frac{\dot {r}}{r}%
\right) ^{2}-3\frac{\dot {r}}{r}\frac{\dot {q}_{1}}{q_{1}}%
-\left( \frac{\dot {q}_{1}}{q_{1}}\right) ^{\mbox{\bf .}}+\frac{1}{2}%
\sum\limits_{\alpha =1}^{3}\left( \frac{\dot {q}_{\alpha }}{q_{\alpha
}}\right) ^{2},
\]
\[
G_{2}^{2}=2\frac{\ddot {r}}{r}+\left( \frac{\dot {r}}{r}%
\right) ^{2}-3\frac{\dot {r}}{r}\frac{\dot {q}_{2}}{q_{2}}%
-\left( \frac{\dot {q}_{2}}{q_{2}}\right) ^{\mbox{\bf .}}+\frac{1}{2}%
\sum\limits_{\alpha =1}^{3}\left( \frac{\dot {q}_{\alpha }}{q_{\alpha
}}\right) ^{2}.
\]
Subtracting from $G_{1}^{1}$ the component $G_{2}^{2}$ one obtains:
\[
3\frac{\dot {r}}{r}\left( \frac{\dot {q}_{2}}{q_{2}}-\frac{%
\dot {q}_{1}}{q_{1}}\right) +\left( \frac{\dot {q}_{2}}{q_{2}}-%
\frac{\dot {q}_{1}}{q_{1}}\right) ^{\mbox{\bf .}}=0.
\]
Entering in the last equation the notification $Q_{\alpha \beta }=\left(
\dot {q}_{2}/q_{2}-\dot {q}_{1}/q_{1}\right) $\ let's have:
\[
3\frac{\dot {r}}{r}+\frac{\dot {Q}_{\alpha \beta }}{Q_{\alpha
\beta }}=0,
\]
that after integration gives:
\[
Q_{\alpha \beta }=C_{\alpha \beta }/r^{3},
\]
where $C_{\alpha \beta }$ are integration constants. From here we get:
\begin{equation}
\label{eq5}
\frac{\dot {q}_{\alpha }}{q_{\alpha }}=\frac{C_{\alpha }}{r^{3}},
\end{equation}
and according to requirements (\ref{eq3}), $\sum\limits_{\alpha =1}^{3}C_{\alpha }=0$%
. Thus integrating the last equation one finds:
\begin{equation}
\label{eq6}
q_{\alpha }=A_{\alpha }\exp \left\{ C_{\alpha }\int \frac{dt}{r^{3}}\right\}
,
\end{equation}
where $A_{\alpha }$ are integration constants and $\prod\limits_{\alpha
=1}^{3}A_{\alpha }~=~1$. Now, using the relation (\ref{eq5}), from (\ref{eq4}) is got:
\begin{eqnarray}
\label{eq7}
-R_{0}^{0}&=&3\frac{\ddot {r}}{r}+\frac{1}{r^{6}}\sum\limits_{\alpha
=1}^{3}C_{\alpha }^{2},\nonumber \\
-R&=&6\left[ \frac{\ddot {r}}{r}%
+\left( \frac{\dot {r}}{r}\right) ^{2}\right] +\frac{1}{r^{6}}%
\sum\limits_{\alpha =1}^{3}C_{\alpha }^{2},
\end{eqnarray}
$\sum\limits_{\alpha =1}^{3}C_{\alpha }^{2}$ determines an anisotropy of the
given model.

\par\bigskip
\noindent{\large \bf 3.\quad Quantum evolution of the Universe}
\par\bigskip

\noindent The possibility of the nucleation of an open Universe circumscribed in
Introduction gives the basis for the consideration of the quantum creation
of a flat Universe. As is known the basic equation of the quantum
cosmology is the Wheeler-DeWitt (WDW) equation. For its making we shall
consider the theory of a scalar field $\varphi $ with Lagrangian
\begin{equation}
\label{eq8}
L=-R/2+\left( \partial _{\mu }\varphi \right) ^{2}/2-V(\varphi )
\end{equation}
or using the expression for scalar curvature $R$ from (\ref{eq4}) one
obtains
\begin{equation}
\label{eq9}
L=-3r\dot{r}^{2}+\frac{r^{3}}{2}\sum\limits_{\alpha =1}^{3}\left( \frac{\dot{%
q}_{\alpha }}{q_{\alpha }}\right) ^{2}+r^{3}\left[ \frac{1}{2}\dot{\varphi}%
^{2}-V(\varphi )\right]
\end{equation}
(accurate to complete derivative with respect to $t$). The relevant conjugate
momentums are equal
\begin{eqnarray}
\label{eq10}
p_{\varphi }&=&\frac{\partial L}{\partial \dot{\varphi}}=r^{3}\dot{\varphi},%
\,\,\,p_{r}=\frac{\partial L}{\partial \dot{r}}=-6r\dot{r}, \nonumber \\%
p_{q_{\alpha }}&=&\frac{\partial L}{\partial \dot{q}_{\alpha }}=r^{3}%
\frac{\dot{q}_{\alpha }}{q_{\alpha }^{2}}
\end{eqnarray}
and the Hamiltonian of the system is
\begin{equation}
\label{eq11}
H=p_{\varphi }\dot{\varphi}+\,\,p_{r}\dot{r}+\,\sum\limits_{\alpha
=1}^{3}\,p_{q_{\alpha }}\dot{q}_{\alpha }-L.
\end{equation}
Let's note, that for deriving the exact equations it was necessary to use
the expression (\ref{eq4}) for scalar curvature instead of (\ref{eq7}).
Using the last one is impossible whereas in this one the integration for
elimination of $\dot{q}_{\alpha }/q_{\alpha }$\ is already yielded. It is
intolerable as actually there is the deletion of variables $q_{\alpha }$ and
thus truncation of the Hamiltonian. Let's note also, that if to use the last
of relations (\ref{eq10}) and expression $\dot{q}_{\alpha }/q_{\alpha
}=C_{\alpha }/r^{3}$\ from (\ref{eq5}) exchanging simultaneously $%
p_{q_{\alpha }}\rightarrow \widehat{p}_{q_{\alpha }}$ (here $\widehat{p}%
_{q_{\alpha }}=-i\partial /\partial q_{\alpha }$), it is easy to obtain that
$q_{\alpha }\widehat{p}_{q_{\alpha }}\Psi =C_{\alpha }\Psi $. The last means
that in our case $\Psi $\ is the eigenfunction of the operator $\widehat{p}%
_{q_{\alpha }}$. It allows with account (\ref{eq5}) to write the Hamiltonian
(\ref{eq11}) as:
\begin{equation}
\label{eq12}
H=\frac{1}{2}\frac{p_{\varphi }^{2}}{r^{3}}-\frac{p_{r}^{2}}{12r}%
+r^{3}V(\varphi )+\left( \sum\limits_{\alpha =1}^{3}C_{\alpha }^{2}\right)
/2r^{3}.
\end{equation}
Quantizing (\ref{eq12}) by replacement of momentums $p_{\varphi }$%
\ and $p_{r}$\ on $-i\partial /\partial \varphi $\ and $-i\partial /\partial
r$\ accordingly and also using the rescaling $\varphi \rightarrow \sqrt{6}%
\Phi $ we obtain the Klein-Gordon equation
\[
\left[ \frac{1}{r^{p}}\frac{\partial }{\partial r}\left( r^{p}\frac{\partial
}{\partial r}\right) -\frac{1}{r^{2}}\frac{\partial ^{2}}{\partial \Phi ^{2}}%
-U_{ef}\right] \psi (r,\Phi )=0,
\]
\begin{equation}
\label{eq13}
U_{ef}=-6\left(
\sum\limits_{\alpha =1}^{3}C_{\alpha }^{2}\right) /r^{2}-12r^{4}V(\Phi ).
\end{equation}
It is the required WDW equation in minisuperspace of the variables $r$ and $%
\Phi $. In this equation the parameter $p$\ represents the ambiguity in the
ordering of noncommuting operators $r$ and $p_{r}$. Let's emphasize, that the
wave function of the Universe $\Psi $\ does not depend on time. This
circumstance is valid for the closed Universe by virtue of equality to zero
of its total energy remains valid and for the flat Universe on the basis of
given in Introduction reasonings.

The transformation of the wave function
\[
\psi =r^{-p/2}\Psi
\]
allows to eliminate first derivative in (\ref{eq13})
\[
\left[ \frac{\partial ^{2}}{\partial r^{2}}-\frac{1}{r^{2}}\frac{\partial
^{2}}{\partial \Phi ^{2}}-U_{ef}\right] \psi (r,\Phi )=0,
\]
\begin{equation}
\label{eq14}
U_{ef}=-%
\frac{p}{2}\left( 1-\frac{p}{2}\right) \frac{1}{r^{2}}-\frac{6
\sum\limits_{\alpha =1}^{3}C_{\alpha }^{2}} {r^{2}}-12r^{4}V(\Phi ).
\end{equation}
Let's note, that in expression for an effective potential energy there is an
addend of \ ''centrifugal energy'' $-p\left( 1-p/2\right) /2r^{2}$.

Is explored Eq. (\ref{eq14}) with a various form of the potential $%
V(\Phi )$\ and by choice of parameter $p$.

\par\bigskip

\noindent \textbf{A. Above barrier reflection of the wave function of the Universe}

\par\bigskip

\noindent 1. {\it de Sitter minisuperspace.} Let's consider the simplest case of
minisuperspace model, when the factor ordering $p=0$ and potential $%
V(\Phi )$\ represents constant vacuum energy $\varepsilon _{v}$ creating an
effective cosmological constant. Then WDW Eq. (\ref{eq14}) will take
the form of the one-dimensional Schr\"{o}dinger equation

\[
\left[ -\frac{d^{2}}{dr^{2}}+U_{ef}\right] \Psi
(r)=0,
\]
\begin{equation}
\label{eq15}
U_{ef}=-6\left( \sum\limits_{\alpha =1}^{3}C_{\alpha
}^{2}\right) /r^{2}-H^{2}r^{4},
\end{equation}
where $H^{2}=12\varepsilon _{v}$\ is the Hubble parameter. Entering $\rho
=H^{1/3}r$\ and $\gamma =6\sum\limits_{\alpha =1}^{3}C_{\alpha }^{2}$\ we
can rewrite (\ref{eq15}) as
\begin{equation}
\label{eq16}
\left[ -\frac{d^{2}}{d\rho ^{2}}+U_{ef}\right] \Psi (\rho
)=0,\,\,\,\,U_{ef}=-\gamma /\rho ^{2}-\rho ^{4}.
\end{equation}
Eq. (\ref{eq16}) describes the motion of a ''particle'' with
zero-point energy in the field of the effective potential $U_{ef}$. The
interesting feature of the given potential is that near to the origin of
coordinates it is approach infinity under the law $U_{ef}\approx -\gamma
/\rho ^{2}$ (that is we can neglect the second term in potential). As is
known [9] , this case is intermediate between when there are
usual stationary states and the cases when happens the ''collapse'' of a
particle in the origin of coordinates. Therefore, it is necessary to carry
out the additional analysis here.

For this purpose we shall search near $\rho =0$ for solution of
\begin{equation}
\label{eq17}
\left[ \frac{d^{2}}{d\rho ^{2}}+\frac{\gamma }{\rho ^{2}}\right] \Psi (\rho
)=0
\end{equation}
as $\Psi \sim \rho ^{s}$ that at substitution in (\ref{eq17}) gives:
\[
s^{2}-s+\gamma =0
\]
with roots

\[
s_{1}=1/2+\sqrt{1/4-\gamma },\,\,\,\,s_{2}=1/2-\sqrt{1/4-\gamma }.
\]
Further, we select around the origin of coordinates small area of radius $%
\rho _{0}$ and is exchanged in it the function $-\gamma /\rho ^{2}$\ by the
constant $-\gamma /\rho _{0}^{2}$. Having defined wave functions in such
''cut off'' field then we shall look what happens at the passage to the
limit $\rho _{0}\rightarrow 0$.

Let's assume at first that $\gamma <1/4$. Then \linebreak $s_{1}~>~s_{2}~>~0$ and at $\rho
>\rho _{0}$ the general solution of Eq. (\ref{eq17}) looks like (at
small $\rho $)
\begin{equation}
\label{eq18}
\Psi =A\rho ^{s_{1}}+B\rho ^{s_{2}}
\end{equation}
( $A,B$ are the constants). At $\rho <\rho _{0}$\ the solution of
\[
\frac{d^{2}\Psi }{d\rho ^{2}}+\frac{\gamma }{\rho ^{2}}\Psi =0,
\]
finiteness in the origin of coordinates looks like
\[
\Psi =C\sin (k\rho ),\,\,\,\,\,k=\sqrt{\gamma }/\rho _{0}.
\]
At $\rho =\rho _{0}$\ function $\Psi $\ and its derivative should be
continuous functions. Therefore, we can write one of requirements as the
requirement of the continuity of the logarithmic derivative with respect to
$\Psi $ that gives in
\[
\sqrt{\gamma }ctg\sqrt{\gamma }=\frac{s_{1}\rho _{0}^{s_{1}-s_{2}}+(B/A)s_{2}%
}{\rho _{0}^{s_{1}-s_{2}}+(B/A)},
\]
or solving rather $(B/A)$ we obtain

\begin{equation}
\label{eq19}
B/A=const\,\rho _{0}^{s_{1}-s_{2}}.
\end{equation}
Passing on to the limit $\rho _{0}\rightarrow 0$ we find that $%
B/A\rightarrow~0$. Thus from two solutions (\ref{eq18}) remains
\begin{equation}
\label{eq20}
\Psi =A\rho ^{s_{1}}.
\end{equation}
Let now $\gamma >1/4$. Then $s_{1}$\ and $s_{2}$\ are complex:
\[
s_{1}=-1/2+i\sqrt{\gamma -1/4},\,\,\,s_{2}=s_{1}^{\ast }.
\]
By analogy to the previous reasonings we again come to equality (\ref{eq19})
which at substitution of values $s_{1}$\ and $s_{2}$ gives
\begin{equation}
\label{eq21}
B/A=const\,\rho _{0}^{i\sqrt{4\gamma -1}}.
\end{equation}
At $\rho _{0}\rightarrow 0$\ this expression does not approach to any limit
so the direct passage to the limit $\rho _{0}\rightarrow 0$\ is impossible.
The general form of the real solution of (\ref{eq17}) can be represented as
following:
\begin{equation}
\label{eq22}
\Psi =const\,\sqrt{\rho }\cos \left( \sqrt{\gamma -1/4}\ln \left( \rho /\rho
_{0}\right) +const\right).
\end{equation}
This function has zeros which number unlimited grows with decreasing of $%
\rho _{0}$. Then at any finite value of the total energy $E$\ the ''normal
state'' of a ''particle'' in given field corresponds to the energy $E=-\infty $.
As ''particle'' is in infinitesimal area around of the origin of coordinates
there is the ''collapse'' of the ''particle'' on centre.

Further, we find the vector of the probability density flux near to zero. In
our one-dimensional case we have:
\[
{\mathbf{j}}=\frac{i}{2}\left( \Psi ^{\ast }\frac{\partial \Psi }{\partial
\rho }-\Psi \frac{\partial \Psi ^{\ast }}{\partial \rho }\right).
\]
It is easy to obtain from here that: 1) in the case $\gamma <1/4$\ at use $%
\Psi $\ from (\ref{eq20}) relevant probability density flux ${\mathbf{j}}=0$
(as well as for any real wave function). 2)~At $\gamma >1/4$\ with the
account $\Psi $\ from (\ref{eq22}) we have $j=\mp \sqrt{\gamma -1/4}$. The
upper sign corresponds to the ingoing and the lower one to the outgoing
wave. The obtained result means that there is the constant probability
density flux near to the origin of coordinates.

Thus we have two types of the behaviour of the wave function of a
''particle'' at various values of the parameter of an anisotropy $\gamma $:
1) at $\gamma <1/4$\ the wave function near to the origin of coordinates
tent to zero; 2) at $\gamma >1/4$\ happens the ''collapse'' of a
''particle'' on centre.

Let's note that the probleme of falling of a particle on centre
has been already studied in isotropic model with the matter equation
of state $p=\epsilon$ in [10].
\begin{figure}
\centerline{\psfig{file=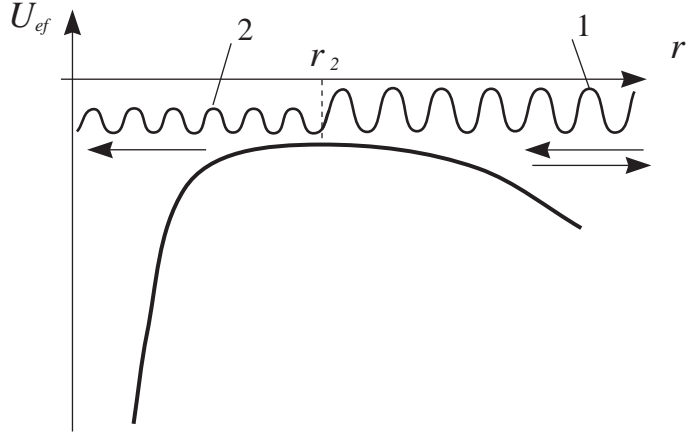}}
\caption{\it{The above barrier reflection of wave function:
1 are ingoing and reflected waves; 2 is the  passed wave.}}
\end{figure}
Let's consider further the motion of a ''particle'' in the semiclassical
approximation. From the form of potential (\ref{eq16}) and equality to zero
of the total energy of the Universe follows that the examination of
Eq. (\ref{eq16}) is reduced to the one-dimensional problem about above
barrier reflection (Fig. 1), i.e. to reflection of a ''particle'' with energy
exceeding height of the barrier. In our case ''particle'' goes down to some
''turning point'' $\rho _{2}$ in which it changes the direction of motion on
an inverse. The given point represents a complex solution of the equation $%
U_{ef}=0$, namely $\rho _{2}=\gamma ^{1/6}\exp \left( i\pi (2n+1)/6\right) $%
. Then the required reflectance $R$ is found as [9]:
\begin{equation}
\label{eq23}
R=\exp \left( -4\,{\rm Im}\int_{\rho _{1}}^{\rho _{2}}\left[ -U_{ef}\right]
^{1/2}\right)
\end{equation}
(here $\rho _{1}$\ is any point on the real axis). In the considered case
using (\ref{eq16}) and (\ref{eq23}) we have:
\begin{equation}
\label{eq24}
R=\exp \left( -\frac{2}{3}\sqrt{\gamma }\pi \right).
\end{equation}

The last expression is obtained with use of the semiclassical approximation.
It is useful also to find the exact solution of Eq. (\ref{eq16})
which looks like:
\begin{equation}
\label{eq25}
\Psi (\rho )=C_{1}\sqrt{\rho }J\left( \frac{1}{3}\sqrt{1/4-\gamma },\frac{1}{%
3}\rho ^{3}\right)
+C_{2}\sqrt{\rho }\,Y\left( \frac{1}{3}\sqrt{1/4-\gamma },%
\frac{1}{3}\rho ^{3}\right),
\end{equation}
where $C_{1}$ and\ $C_{2}$\ are integration constants, $J(\nu ,z)$, $Y(\nu
,z)$ - the Bessel functions of the first and second kind accordingly, $%
\nu ,z$\ are index and argument of these functions. For the finding of the
reflectance it is necessary to consider the behaviour of this solution on
major distances from the origin of coordinates. In this case $\Psi (\rho )$\
describes ingoing and reflected waves and the reflectance will be equal to
the ratio of amplitudes of these waves. The asymptotic form of the function $%
\Psi (\rho )$\ at $\rho \rightarrow \infty $\ is
\begin{equation}
\label{eq26}
\Psi (\rho )\approx \sqrt{\frac{6}{\pi \rho ^{2}}}\cos \left( \frac{1}{3}%
\rho ^{3}-\frac{1}{6}\pi \sqrt{1/4-\gamma }-\frac{\pi }{4}\right) .
\end{equation}
It is clear from here that:
\[
\mbox {for}\,\,\,\gamma <1/4:\,\,\,R=1
\]
\begin{equation}
\label{eq27}
\mbox {for}\,\,\,\gamma >1/4:\,\,\,
R=\exp \left( -\frac{2}{3}\pi \sqrt{\gamma -1/4}\right) .
\end{equation}
The second expression from (\ref{eq27}) at $\gamma \gg 1/4$\ coincides with
reflectance from the semiclassical approximation (\ref{eq24}).

Being based on results obtained at the analysis of behaviour of the wave
function of the Universe near to the origin of coordinates is concluded that
at $\gamma <1/4$\ does not happen the accumulation of the wave function at $%
\rho \rightarrow 0$\ and consequently takes place the complete reflection of
the wave function from the barrier ($R=1$). In the case $\gamma >1/4$ which
corresponds to collapse of a ''particle'' on centre the reflectance $R$\
becomes less than 1. It happens because there is nonzero probability density
flux in the infinitesimal area around of the origin of coordinates. Let's
note that at the approach to zero the problem, generally speaking, ceases to
be stationary. It gives that the wave function can accumulate in this area.

\par\medskip

\noindent 2. \vspace{1pt}{\it Variable scalar field.} Let's consider further the
case of Eq. (\ref{eq14}) when the factor ordering $p$\ is equal to zero as
before, but potential of the scalar field is variable: $V(\varphi
)=m^{2}\varphi ^{2}/2$. Introducing the rescaling $\varphi \rightarrow \sqrt{%
6}\Phi $ and $m\rightarrow \mu /6$ we shall write the WDW equation as:
\[
\left[ -\frac{\partial ^{2}}{\partial r^{2}}+\frac{1}{r^{2}}\frac{\partial
^{2}}{\partial \Phi ^{2}}+U_{ef}\right] \Psi =0,
\]
\begin{equation}
\label{eq28}
U_{ef}=-\frac{%
\gamma }{r^{2}}-\mu ^{2}r^{4}\Phi ^{2}.
\end{equation}
The finding of the analytical solution of the obtained equation represents a
complex problem. Therefore, for its examination we shall search the WKB
solution as $\Psi _{c}=e^{-iS}$. The relevant equation for action $S(r,\Phi )
$\ is
\begin{equation}
\label{eq29}
-\left( \frac{\partial S}{\partial r}\right) ^{2}+\frac{1}{r^{2}}\left(
\frac{\partial S}{\partial \Phi }\right) ^{2}-U_{ef}=0.
\end{equation}
For finding of the solution of this nonlinear differential equation it is
possible to reduce it to system of the ordinary differential equations
called the characteristic system of the given partial equation. Using this
system it is possible to construct an integrated surface of Eq.
(\ref{eq29}), consisting from the characteristics. The required system of the
characteristics written with respect to arbitrary parameter $t$\ has a form
[11] :
\begin{eqnarray}
\label{eq30}
r^{\prime }(t)&=&-2p,\,\,\,\Phi ^{\prime }(t)=\frac{2}{r^{2}}q, \nonumber \\
S^{\prime }(t)&=&-2\left( \frac{\gamma }{r^{2}}+\mu ^{2}r^{4}\Phi
^{2}\right), \nonumber \\
p^{\prime }(t)&=&\frac{2}{r^{3}}q^{2}+2\frac{\gamma }{r^{3}}-4\mu
^{2}r^{3}\Phi ^{2}, \\
q^{\prime }(t)&=&-2\mu ^{2}r^{4}\Phi, \nonumber
\end{eqnarray}
where $^{\prime }$ denote a derivative with respect to $t$ and introducing denotations $%
p=\partial S/\partial r $, $q=\partial S/\partial \Phi $. The obtained system
of equations describes an one-dimensional motion of a\ ''particle'' along
the characteristic. In this case monotonically varying parameter $t$ can
play a role of time [12]\ in due course there is an evolution
of the Universe.

The obtained system of characteristics can be use for finding of dependence
of coefficient of above barrier reflection $R$\ of wave function of the
Universe from value of the field $\Phi $.

But before we shall note one useful possibility of simplification of making
of a calculation. Expand an effective potential $U_{ef}$\ from (\ref{eq16}) near to
a maximum in a Taylor series:
\begin{equation}
\label{eq31}
U_{ef}=U_{\max }+\frac{\alpha }{2}\left( \rho -\rho _{\max }\right) ^{2},
\end{equation}
here $U_{\max }$\ and $\rho _{\max }$\ are a maximal value of potential and
relevant value of $\rho $, and $\alpha $ \ is a value of a second derivative with
respect to $U_{ef}$ \ from (\ref{eq16}) in the point $\rho _{\max }$. The values of
specified quantities are easily finding from (\ref{eq16}):
\[
U_{\max }=-3\left(\frac{\gamma}{2}\right)^{2/3},\,\,
\rho _{\max }=\left(\frac{\gamma}{2}\right)^{1/6}
\]
\[
\alpha =-24\left(\frac{\gamma}{2}\right)^{1/3}.
\]
Then using (\ref{eq23}) and (\ref{eq31}) we have
\begin{equation}
\label{eq32}
R=\exp \left( -\frac{1}{2}\sqrt{\frac{3}{2}}\sqrt{\gamma }\pi \right) ,
\end{equation}
that approximately coincides with (\ref{eq24}). Thus in an one-dimensional case
there is an possibility to use approximate expression (\ref{eq31}) instead of exact one.
\begin{figure}
\centerline{\psfig{file=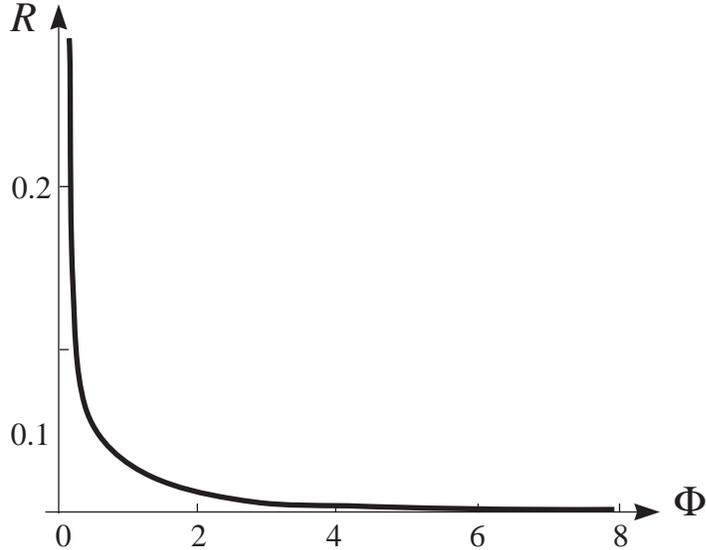}}
\caption{\it{Dependence of coefficient of above barrier reflection $R$ from value
        of the scalar field $\Phi$.}}
\end{figure}

Further, we shall search for the solution of system (\ref{eq30}) numerically. Using
(\ref{eq23}) and (\ref{eq31}) it is possible to obtain the following form of dependence $%
R(\Phi )$\ (Fig. 2). It is obvious that the reflectivity promptly decreases
with increasing of value of the field $\Phi $. After reflection of the wave
function the following necessary stage of evolution of the Universe is the
inflationary period providing the ''stretch'' of linear size of the Universe
with Planck up to macroscopic. For ensuring of the sufficiently long period
of inflation it is necessary the realization of two requirements on scalar
field (see e.g. [13] ): 1) it should be a Planck order;
2) its values should vary slowly with time. The realization first of these
requirements corresponds rather small, but nevertheless distinct from zero
reflectivity $R$\ (see Fig. 2). Solving system (\ref{eq30}) will be easily convinced
that is fulfilled second of the posed requirements also.

\par\bigskip

\noindent \textbf{B. The wave function bouncing off a potential wall}

\par\bigskip

\noindent Now we shall consider Eq. (\ref{eq14}) in a case when the parameter $p$ is
different from zero and there is a potential of the scalar field $V(\varphi
)=m^{2}\varphi ^{2}/2$. At such setting of a problem we have two essentially
various variant of the effective potential $U_{ef}$. At realization of a
requirement $0\leq p\leq 2$\ the form of $U_{ef}$\ from (\ref{eq14}) as a matter of
fact by nothing differs from the case when $p=0$\ and the relevant
consideration will be made by analogy to the problem studied in the previous
section. In a case omissions of the mentioned above requirement the
qualitatively new statement of a problem is possible. For this purpose the
realization of one requirement is necessary: $\left| -p^{2}/2+p/2\right|
>\gamma $,\ which ensures occurrence in the effective potential of a
positive factor before $1/r^{2}$\ greater than $\gamma $.
Thus there is a possibility of occurrence of a repelling potential wall.
Then the effective potential in (\ref{eq14}) will be:
\begin{equation}
\label{eq33}
U_{ef}=\frac{\left| -p^{2}/2+p/2\right| }{r^{2}}-\frac{\gamma }{r^{2}}-\mu
^{2}r^{4}\Phi ^{2}-\varepsilon .
\end{equation}
Thus the influence of a massless scalar field (e. g. photon gas) with energy
density $\varepsilon $ also is taken into account, the sense of which
introduction will be explained below. The form of potential is shown on
Fig.3. Then Eq. (\ref{eq14}) describes a motion of \ a ''particle'' with a
zero-point energy in potential $U_{ef}(r,\Phi )$.
\begin{figure}
\centerline{\psfig{file=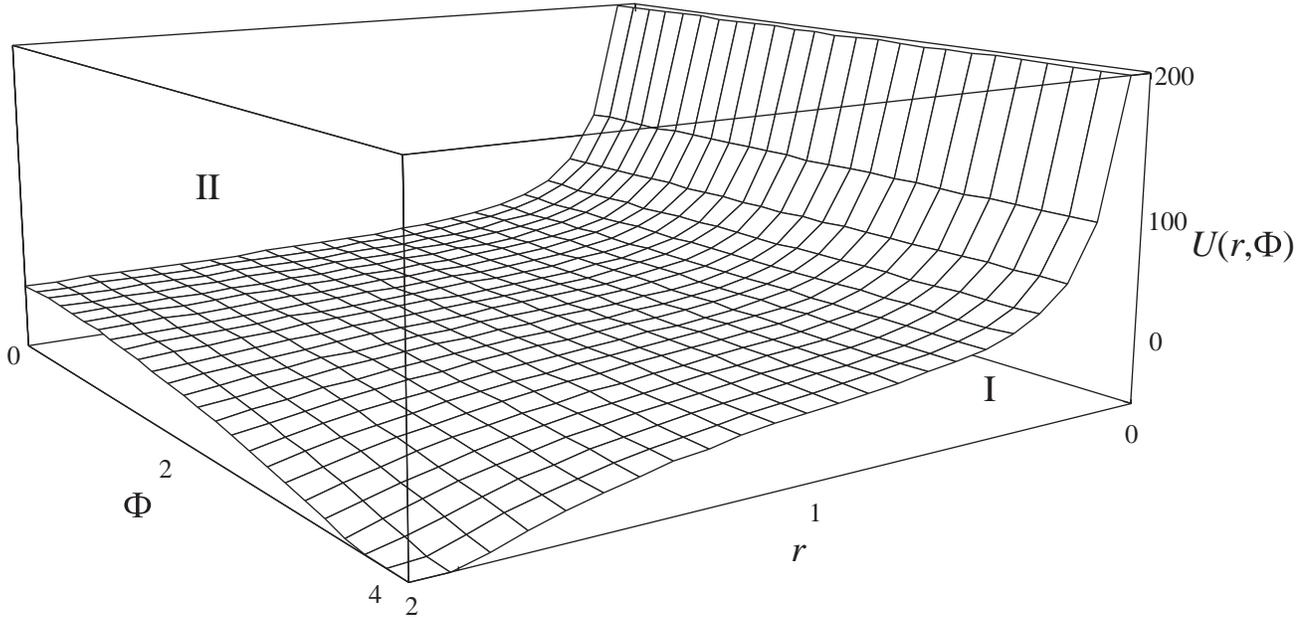}}
\caption{\it{The effective potential $U_{ef}$. I is a classically forbidden area,
         II is a classically allowed area.}}
\end{figure}

As is obvious from Fig.3 the effective potential parts space on two areas:
classically forbidden (interior) and classically allowed (exterior). At
transition of a\ ''particle'' in the I-st area  its wave function will be
exponentially decay (as $U_{ef}(r,\Phi )$\ approach infinity at $%
r\rightarrow \infty $) and the relevant probability of realization of the
given state tends to zero.

On the other hand, the wave function $\Psi $\ describes the wave incident on
the barrier $U_{ef}(r,\Phi )$ on the part of major $r$ and the wave reflex
from the barrier. The physical sense of the given statement consists that
the incident wave describes the contracting Universe and reflex - expanding one.

We shall consider an evolution of the scalar field $\Phi $\ at stages of
contraction and expansion. For this purpose we shall search the WKB
solution of the equation (\ref{eq14}) with potential (\ref{eq33}) in classically allowed
range as $\Psi _{c}~=~e^{-iS}$. The relevant equation for action $S(r,\Phi )$\
will be:
\begin{equation}
\label{eq34}
-\left( \frac{\partial S}{\partial r}\right) ^{2}+\frac{1}{r^{2}}\left(
\frac{\partial S}{\partial \Phi }\right) ^{2}-U_{ef}=0.
\end{equation}
By analogy to the case of Eq. (\ref{eq29}) we shall make system of the
characteristics with respect to arbitrary parameter $t$:
\begin{eqnarray}
\label{eq35}
r^{\prime }(t)&=&-2p,\,\,\,\Phi ^{\prime }(t)=\frac{2}{r^{2}}%
q, \nonumber \\
S^{\prime }(t)&=&2\left( \frac{\left| -p^{2}/2+p/2\right| }{r^{2}}-%
\frac{\gamma }{r^{2}}-\mu ^{2}r^{4}\Phi ^{2}-\varepsilon \right), \nonumber \\
p^{\prime }(t)&=&\frac{2}{r^{3}}q^{2}-2\frac{\left| -p^{2}/2+p/2\right| }{%
r^{3}}+2\frac{\gamma }{r^{3}}-4\mu ^{2}r^{3}\Phi ^{2}, \\
q^{\prime}(t)&=&-2\mu ^{2}r^{4}\Phi. \nonumber
\end{eqnarray}
As well as in case of Eq. (\ref{eq29}) system of equations (\ref{eq35}) is
describes an one-dimensional motion of a ''particle'' along the
characteristic. Thus the behaviour of\ a\ ''particle'' is similar on bounce
off a potential barrier $\Phi (r)$  made at cross of the effective
potential  with the plane $\Phi -r$\ (that is when $U_{ef}=0$). Therefore, a
semiclassical wave function $\Psi _{c}$\ describes an ensemble of classical
universes evolving along the characteristics $S$. Then the ensemble of these
characteristics can be considered as the trajectories of a motion with the
various initial conditions.

We shall note here that introduced earlier the massless\ scalar field with
the energy density $\varepsilon $ is necessary that the Universe having a
zero total energy always remained in classically allowed area (Fig. 4).
\begin{figure}
\begin{picture}(50,160)
\put(-230,-100){\includegraphics{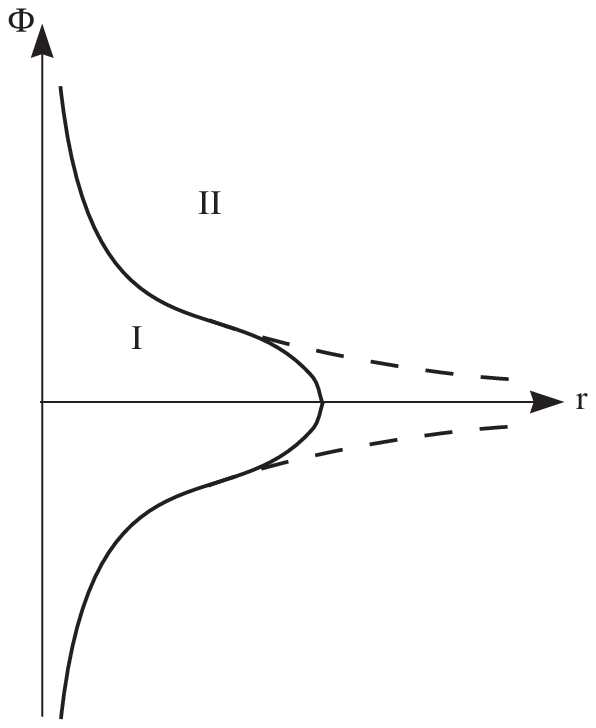}}
\put(-5,-55){\parbox{5cm}{Figure 4. \it{Conventional form of the
potential barrier with $\varepsilon$. I is a classically forbidden
area; II is a classically allowed area. The dashed lines - form of
the barrier without $\varepsilon$.} }}

\put(20,-100){\includegraphics{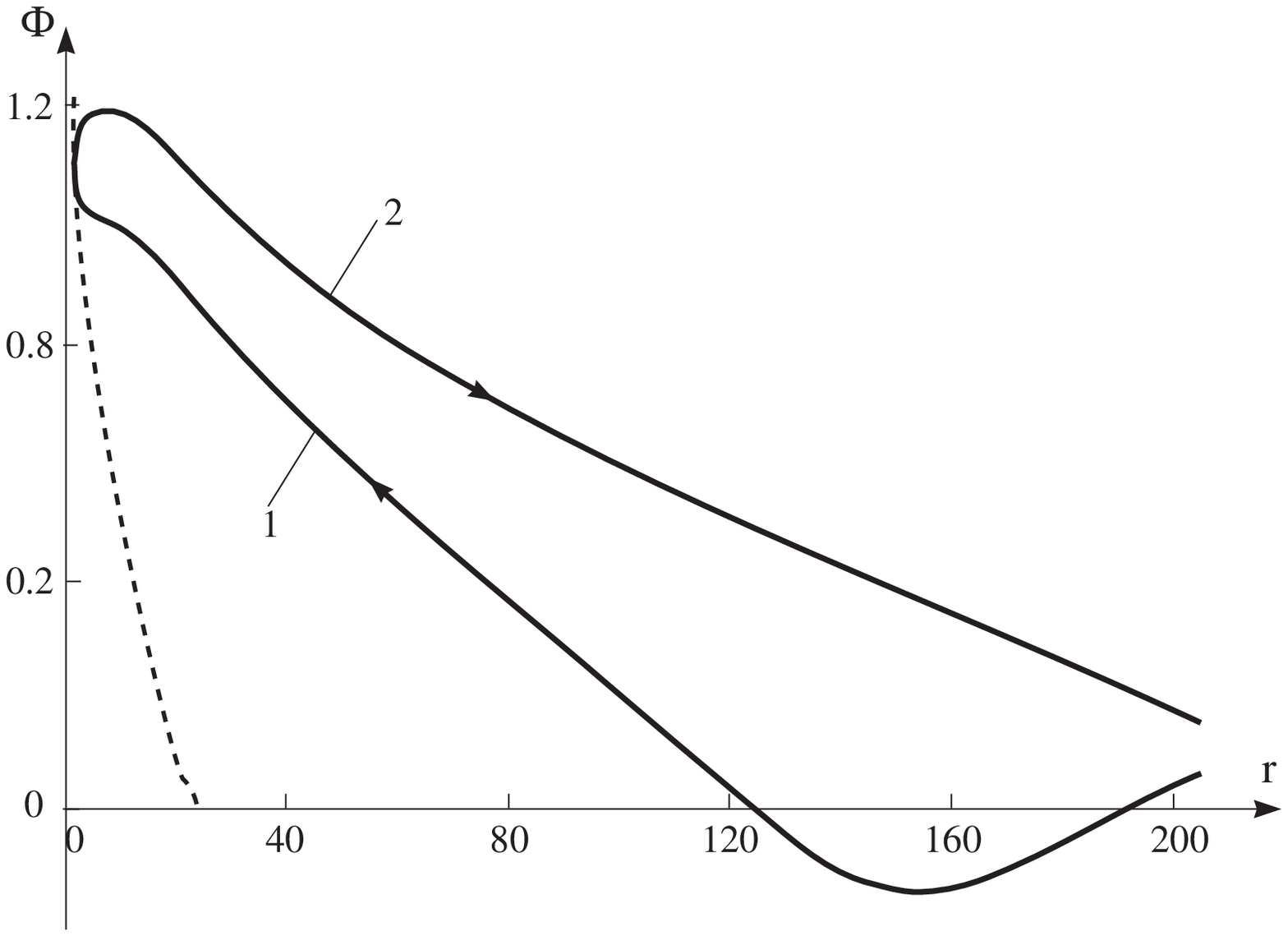}}
\put(160,-40){\parbox{9.5cm}{Figure 5. \it{The diagram of
dependence of the scalar field $\Phi$ from $r$. 1 is an evolution
of the field on the contracting stage, 2 - during the inflation.
The dashed line conventionally shows the form of the potential
barrier.} }}
\end{picture}
\end{figure}

As it was specified above the evolution of the Universe is described by two
stages: at first regime of contracting and then, after bounce off, regime
of expansion. At the stage of contracting the field makes oscillations with
increasing amplitude. After bounce off two variants are possible: 1) the
field $\Phi $\ amount to rather major value and after reflection varies
slowly that corresponds to an inflationary period and further transfers on
scalaron stage (Fig. 5); 2) $\Phi $\ hasn't  amount to major values and
having reflected at once makes fast oscillations losing the energy. Thus,
naturally, there is no inflationary stage (or it is too short).

\par\bigskip
$\phantom{x}$
\par\bigskip
\par\bigskip
\par\bigskip
\par\bigskip
\par\bigskip
\par\bigskip
\par\bigskip
It is necessary apart to note that in considered model the presence of a
repelling potential wall does not allow the Universe to collapse.

\par\bigskip

\noindent {\large \bf 4.\quad Conclusions}

\par\bigskip

\noindent In the submitted paper we have considered an anisotro\-pic cosmological
Bianchi type I model with the scalar field. Distinctive feature at the
solution of the given problem was representation of three scale factors in
such manner that the final Einstein's equations have turned out dependent
only from one function $r(t)$.

In the quantum approach the basic equation of the quantum cosmology -~WDW
equation (\ref{eq13}) was obtained. The examination of the latter was
reduced to considering the following problems:

1. The simplest model with constant scalar field playing a role of an
effective cosmological constant was studied. It has allowed to reduce the WDW
equation to the one-dimensional Schr\"{o}dinger equation. The presence of
the parameter of an anisotropy $\gamma $\ in the effective potential gives
in interesting feature: there is some critical value of this parameter at
which there is partitioning the problem into two variants. In the first case
at $\gamma >1/4$ the collapse of the wave function on centre takes place.
Thus there is the constant nonzero probability density flux $j$ that means
an possibility of accumulation of the wave function in close to the origin
of coordinates area. In the second case ($\gamma <1/4$) the collapse of the
''particle'' on centre misses and $j=0$.

Further, the problem of finding of coefficient of above barrier reflection $%
R $ from the cosmological singularity of the wave function of the Universe $%
\Psi $\ was solved. For this purpose two approaches were used:
1)~semiclassical approximation and 2) finding $R$ as the ratio of amplitudes of
wave functions of reflected and ingoing waves on infinity. Both approaches
give identical results at $\gamma \gg 1/4$. Thus $R<1$\ that means partial
penetration of the wave function into close to zero area and its further
collapse on centre. In case of $\gamma <1/4$\ the second approach gives $R=1$%
\ as against semiclassical one. It speaks about an inapplicability the latter in
such situation and about necessity to use asymptoticses of the exact
solutions.

Let's note that on coefficient of above barrier reflection the influence of
Hubble parameter has not an effect in any way because it is only
renormalaze the scale factor.

2. The model with a variable scalar field with potential $V(\varphi
)=m^{2}\varphi ^{2}/2$ was considered. In view of complexity of finding of
an exact analytical solution the given problem was explored numerically.
With this purpose the semiclassical solution of Eq. (\ref{eq28}) was
found, therefore Eq. (\ref{eq29}) is obtained for which the system of
the characteristics (\ref{eq30}) was obtained. The last one was used for finding of
relationship of the coefficient of the above barrier reflection $R$\ from
value of the field $\Phi $. The obtained relationship shows that $R$\ is
rather great at Planck field. It means that after reflection there is an
possibility of an output of the Universe on a rather long inflationary stage
providing increase of its size with Planck up to macroscopic and a further
output on the standard cosmological scenario.

3. In case of nonzero factor ordering there is essentially new
possibility of a ''bounce off'' a wave function of the Universe from a
repelling potential wall ensured with a form of effective potential (\ref{eq33}). By
analogy to the previous case the solution of Eq. (\ref{eq34}) was found
with the help of system of the characteristics (\ref{eq35}). The obtained results
show that the Universe beginning the evolution with small initial value of $%
\Phi $ at the stage of contraction gathers a field up to Planck. After
bounce off the field will increase still and further Universe transfers to
an inflationary stage (see Fig. 5).

Note that in considered cases above barrier reflection and bounce of a wave
function of the Universe feature process of the quantum creation of the Universe.
The further evolution of the model represents a stage of expansion with prompt
losses of an anisotropy and transition into Friedmann Universe.

\par\bigskip

{\bf Acknowledgements}
\par\bigskip

\noindent We are grateful to A.A. Starobinsky and I.B. Khriplovich for useful discussions
of results.

This work was supported by the research grant~KR-154 of International
Science and Technology Centre (ISTC).

\end{document}